\documentclass[a4paper,11pt]{article}%
\usepackage{geometry}
\usepackage{amsmath}
\usepackage{amsfonts}
\usepackage{amssymb}
\usepackage{graphicx}
\usepackage{indentfirst}
\usepackage{bbold}
\usepackage[small,bf]{caption}
\usepackage{slashed}
\usepackage{xcolor}
\usepackage{subcaption}
\usepackage{cite}
\usepackage[toc,title,page]{appendix}
\numberwithin{equation}{section}

\usepackage{multirow}

\usepackage[normalem]{ulem}
\newcommand\redsout{\bgroup\markoverwith{\textcolor{red}{\rule[0.4ex]{3pt}{0.7pt}}}\ULon}

\newcommand{\PR}[1]{\ensuremath{\left[#1\right]}}
\newcommand{\PC}[1]{\ensuremath{\left(#1\right)}}

\newcommand{\ii}{\mathrm{i}}

\providecommand{\U}[1]{\protect\rule{.1in}{.1in}}

\DeclareMathOperator{\Tr}{Tr}
\usepackage{color}
\definecolor{darkgreen}{rgb}{0,0.35,0}
\definecolor{Rood}{rgb}{1, 0, 0}

\begin{document}

\date{}
\title{ \textbf{Implications of the topological Chern-Simons mass in the Gap Equation
}
}
\author{
{\bf Caroline P.~Felix}$^{a}$\thanks{felix@cycu.edu.tw}\,\, ,
{\bf Chung Wen Kao}$^{a}$\thanks{cwkao@cycu.edu.tw }\,\,,
\\[2mm]
{\small $^a$ \it  Department of Physics and Center for High Energy Physics, }\\
{\small \it  Chung-Yuan Christian University, Chung-Li 32023, Taiwan}\\
 
 
 }

\maketitle

\begin{abstract}
In this paper, we solve the gap equation of the Yang-Mills-Gribov-Zwanziger-Chern-Simon theory by considering the first order in the Chern-Simon topological mass term, $M$. As a result, we find three possible solutions to the gap equation, i.e. three different Gribov parameters. In addition, we analyze the regime of the theory for each of these Gribov parameters and we obtain a different result from the literature.
\end{abstract}

\section{Introduction}
It is well known that the non-perturbative non-Abelian gauge field theory is plagued by gauge copies -- the famous Gribov copies \cite{Gribov:1977wm}. Gribov realized the standard gauge fixing
procedure by Faddeev-Popov is not sufficient to remove all equivalent gauge field
configurations in non-perturbative non-Abelian gauge field theory. To solve this problem, Gribov found a region that is free from these kinds of copies, which is called the Gribov Region. 
In 44 years, a lot of research has been focused on the Gribov problem, see \cite{Singer:1978dk, Dudal:2008sp, Vandersickel:2012tz, Capri:2014jqa, Capri:2015nzw, Dudal:2019ing, Gotsman:2020ryd} for examples. The pioneer work was done by Zwanziger, who figured out how to implement the Gribov region in the action, hence a new formalism called Gribov-Zwanziger formalism was created \cite{Zwanziger:1988jt, Zwanziger:1989mf, Zwanziger:1992qr}, what some authors have used to understand and explain confinement/deconfinement phase transition \cite{Canfora:2013zza,Ferreira_2020,Dudal:2017jfw,Maelger:2018vow}.

Another method to study confinement/deconfinement  is through $3d$ Euclidean Yang-Mills-Chern-Simons theory \cite{Deser:1982vy, Deser:1981wh}. This method is interesting because the topological  Chern-Simons (CS) mass term gives the gluon field an extra mass generating a confinement/deconfinement transition phase in $3d$ Euclidean Yang-Mills \cite{Canfora:2013zza, Ferreira_2020} . 

As Gribov-Zwanziger theory and Yang-Mills-Chern-Simons theory give us  features about the regime of the theory, 
it is intriguing to assemble both theories
and then to solve
the gap equation for the Gribov parameter 
\cite{Gribov:1977wm, Zwanziger:1988jt, Zwanziger:1989mf, Zwanziger:1992qr}
and to analyze the confinement/deconfinement phase transition
in the presence of the CS topological mass $M$ at least in the first order.



The paper is organized as follows: in Section \ref{propagatorsec}, we briefly review the gluon propagator calculus and the regime of the theory before solving the gap equation. In Section \ref{gapeq}, the gap equation for the Gribov parameter is solved. In the Section \ref{propagatorPoles}, we analyze the regime of the theory by using the Gribov parameters found in Section \ref{gapeq}. Finally, we present our conclusions in Section \ref{conclusion}.

\section{The Yang-Mills-Gribov-Zwanziger-Chern-Simons gluon propagator}\label{propagatorsec}

The starting point of our investigation is the local Gribov-Zwanziger action in linear covariant gauge in $3d$ dimensions Enclidean space \cite{Vandersickel:2012tz},
\begin{equation}
\begin{split}
    S ~=~ &\int d^3x \frac{1}{4}\left(F_{\mu\nu}^a\right)^2+\int d^{3}x  \left(\frac{\alpha}{2}b^ab^a+\ii
b^{a}\,\partial_{\mu}A^{a}_{\mu}
+\bar{c}^{a}\partial_{\mu}D^{ab}_{\mu}c^{b}     \right)\\
&+\int d^{3}x\left[\bar{\varphi}^{ac}_{\mu}\partial_{\nu}D^{ab}_{\nu}\varphi^{bc}_{\mu}-\bar{\omega}^{ac}_{\mu}\partial_{\nu}(D^{ab}_{\nu}\omega^{bc}_{\mu})
-g(\partial_{\nu}\bar{\omega}^{an}_{\mu})f^{abc}D^{bm}_{\nu}c^{m}\varphi^{cn}_{\mu}\right]
\\
&-\gamma^{2}g\int d^3 x\left[\vphantom{\frac{1}{2}}f^{abc}A^{a}_{\mu}\varphi^{bc}_{\mu}+f^{abc}A_{\mu}^{a}\bar{\varphi}_{\mu}^{bc}+\frac{3}{g}(N_{c}^{2}-1)\gamma^{2}\vphantom{\frac{1}{2}}\right],
\label{GZ_action}
\end{split}
\end{equation}
with 
$\gamma$\footnote{$\gamma^4=\frac{2\beta Ng^2}{3V(N^2-1)}$} is the Gribov parameter; $F^{a}_{\mu\nu}$ is the field strength tensor, which is defined by the equation:
\begin{equation}
F^{a}_{\mu\nu} ~=~ \partial_{\mu}A^{a}_{\nu} - \partial_{\nu}A^{a}_{\mu} + gf^{abc}A^{b}_{\mu}A^{c}_{\nu}\;; \label{fstr}
\end{equation}
$(\phi,\bar{\phi})$ is a pair of complex-conjugate bosonic fields; $(\omega,\bar{\omega})$ are  anti-commuting complex-conjugate fields;
the fields $({\bar c}^a, c^a)$ are the Faddeev-Popov ghosts; $\alpha$ is the gauge parameter, which is zero for the Landau
gauge, $\partial_{\mu}A_{\mu} = 0$; $b^a$ accounts for the Lagrange
multiplier implementing the gauge condition;
and $D^{ab}_\mu =( \delta^{ab}\partial_\mu + g
f^{acb}A^{c}_{\mu})$ is the covariant derivative in the adjoint representation of $SU(N)$.

Now, by coupling the Chern-Simons action,
\begin{equation}
S_{\text{CS}} ~=~ -\frac{\ii M}{2}\epsilon^{\mu\nu\lambda}
\int d^3x
 \left(A_\mu^a\partial_\lambda
A_\nu^a-\frac{2}{3!}gf^{abc}A_\mu^aA_\nu^bA_\lambda^c\right) \,,
\label{CS_action}
\end{equation}
to \eqref{GZ_action}, we obtain the Yang-Mills-Gribov-Zwanziger-Chern-Simons (YMGZCS) action:
\begin{equation}
\begin{split}
    S_{YMGZCS} ~=~ &\int d^3x \left[\frac{1}{4}\left(F_{\mu\nu}^a\right)^2-\frac{iM}{2}\epsilon_{\mu\nu\lambda}\left(A_\nu^a\partial_\lambda A_\mu^a-\frac{2}{3!}gf^{abc}A_\mu^aA_\nu^bA_\lambda^c\right)\right]\\
    &+\int d^{3}x  \left(\frac{\alpha}{2}b^ab^a+\ii
b^{a}\,\partial_{\mu}A^{a}_{\mu}
+\bar{c}^{a}\partial_{\mu}D^{ab}_{\mu}c^{b}     \right)\\
&+\int d^{3}x\left[\bar{\varphi}^{ac}_{\mu}\partial_{\nu}D^{ab}_{\nu}\varphi^{bc}_{\mu}-\bar{\omega}^{ac}_{\mu}\partial_{\nu}(D^{ab}_{\nu}\omega^{bc}_{\mu})
-g(\partial_{\nu}\bar{\omega}^{an}_{\mu})f^{abc}D^{bm}_{\nu}c^{m}\varphi^{cn}_{\mu}\right]
\\
&-\gamma^{2}g\int d^3 x\left[\vphantom{\frac{1}{2}}f^{abc}A^{a}_{\mu}\varphi^{bc}_{\mu}+f^{abc}A_{\mu}^{a}\bar{\varphi}_{\mu}^{bc}+\frac{3}{g}(N_{c}^{2}-1)\gamma^{2}\vphantom{\frac{1}{2}}\right] .
\label{ymcfjaction}
\end{split}
\end{equation}

The gluon propagator poles in Yang-Milss-Chern-Simons theory in the presence of Gribov ambiguity have already been analyzed in the literature
\cite{Canfora:2013zza}. However, in \cite{Canfora:2013zza}, the authors considered only the zero order of the gap equation expansion in the CS mass term.
Despite the zero order being the most dominant term, we will see that the first order in the CS mass has notable physical implications.


To calculate the gluon propagator from this theory, it is necessary to take only the quadratic part in the gauge field of action \eqref{ymcfjaction}
and integrate it out. Following these steps, one should {end} up with
\begin{equation}
S ~=~ \int \frac{d^3k}{(2\pi)^3}\left(-\frac{1}{2}\tilde{A}_\mu^a(k)Q_{\mu\nu}^{ab}\tilde{A}_\nu^b(-k)\right)
\end{equation}
 where 
 \begin{equation}
 Q_{\mu\nu}^{ab} ~=~ \delta^{ab}\left(\frac{k^4+\gamma^4}{k^2}\delta_{\mu\nu}+\left(\frac{1}{\alpha}-1\right)k_\mu k_\nu +M \epsilon_{\mu\nu\lambda}k_\lambda\right)
 \label{q}
 \end{equation}
and $\gamma^4$ is the Gribov parameter.
In order to obtain the propagator we 
have to
compute the inverse of \eqref{q}, which can be obtained through the following expression,
 \begin{equation}
 Q_{\mu\nu}^{ab}(Q_{\nu\delta}^{bc})^{-1} ~=~ \delta^{ac}\delta_{\mu\delta} \,.
 \label{inversecondition}
 \end{equation}
{The ansatz for the inverse of \eqref{q} reads}
\begin{equation}
(Q_{\nu\delta}^{bc})^{-1} ~=~ \delta^{bc}\left(
F(k)\delta_{\nu\delta}+
B(k) \frac{k_\nu k_\delta}{k^{2}}+
C(k)M\,\epsilon_{\delta\nu\alpha} \frac{ k_\alpha}{k^{2}}\right) \,,
 \label{ansatz}
 \end{equation}
where the coefficients are dimensionless.

The Landau gauge is recovered in the limit $\alpha\to0$, and 
the propagator reads
\begin{eqnarray}
\langle A^a_\mu(k) A^b_\nu(-k)\rangle &=& \delta^{ab}{F(k)}
\left[\left(\delta_{\mu\nu}-\frac{k_\mu k_\nu}{k^2}\right)-
\frac{k^4}{(k^4+\gamma^4)}M\,\epsilon_{\mu\nu\alpha} \frac{k_\alpha}{k^{2}}
\right]  \,.
\label{gribovcspropagator}
\end{eqnarray}
The overall factor
 $F(k)$ is given by
\begin{equation}
F(k)=\frac{(k^4+\gamma^4)k^2}{(k^4+\gamma^4)^2+k^6M^2}  \,.
\label{poles}
\end{equation}

As it is pointed out in \cite{Canfora:2013zza}, the poles of the propagator \eqref{gribovcspropagator} are found by determining the roots of the following
polynomial:
\begin{eqnarray}
P(k^2)&=&(k^4+\gamma^4)^2+k^6M^2\nonumber\\
&=&(k^2+m_1^2)(k^2+m_2^2)(k^2+m_3^2)(k^2+m_4^2) 
\label{polynomialequation}
\end{eqnarray}
where $m_i$ stands for the solutions of the polynomial $P(k^2)$.
The discriminant of $P(k^2)$ is
\begin{equation}
    \Delta_p=256M^4\gamma^{20}-27M^8\gamma^{16}.\label{DeltaP}
\end{equation}
As a result, there are four complex roots for $P(k^2)$, if $\Delta_p>0$ or $\gamma>\sqrt[4]{27} M/4$, and there are two complex and two real roots for $P(k^2)$, if $\Delta_p<0$ or $\gamma<\sqrt[4]{27} M/4$. 

In the study of \cite{Canfora:2013zza}, $\gamma$ (there, it is called $G$) is a function only of the coupling constant $g$, since they have not considered higher orders in the CS mass term $M$. Hence, they can define the regime of the theory or the confinement/deconfinement phase transition by comparing $g$ and $M$. In our case, we cannot do it so fast, since $\gamma$ depends on $g$ and $M$ and the gap equation has more than one solution as we will see in the next section. Then, let us calculate, first, all possible solutions for the gap equation, i.e. let us find $\gamma$'s in Section \ref{gapeq}, and then come back to the regime of the theory analysis in Section \ref{propagatorPoles}. 

\section{Three solutions for the gap equation}
\label{gapeq}


{{In this section, we analyze the contribution of the Chern-Simon mass term to the gap equation.}} 
The gap equation is a self consistent condition obtained through the saddle-point approximation, 
which becomes exact in the thermodynamic limit \cite{Gribov:1977wm,Zwanziger:1988jt,Zwanziger:1989mf,Zwanziger:1992qr}. In other words, the gap equation can be obtained by taking the first derivative of the vacuum energy density,  $\mathcal{E}_{v} $, with respect to $\beta$, computed at the specific value $\beta^*$ that minimizes $\mathcal{E}_{v} $. The vacuum energy density,  $\mathcal{E}_{v} $, is given by \cite{Canfora:2013zza}
\begin{eqnarray}
-V\mathcal{E}_{v}= 
\beta - \ln\beta
-\frac{1}{2} \Tr\ln Q_{\mu\nu}^{ab},
\label{kdjf}
\end{eqnarray}
Then, taking into account the saddle-point in the thermodynamic limit by holding $\gamma^4=\frac{2\beta Ng^2}{3V(N^2-1)}$ finite, we find 
\begin{eqnarray}
&&
\int \frac{d^{3}k}{(2\pi)^{3}}\;
\frac{k^{4} + \gamma_*^4}{\left( k^{4} + \gamma_*^4 \right)^{2} 
+k^{6} M^2}
~=~ 
\frac{3}{2Ng^2}
\,.
\label{gpequation3d}
\end{eqnarray}
The calculus of \eqref{gpequation3d}  is similar to what has been done in \cite{Ferreira_2020}. By solving this integral, step-by-step calculation is found in Appendix \ref{SolGapEg}, we find the following equation:
\begin{equation}
      \gamma^3-\frac{Ng^2}{6\sqrt{2} \pi}\gamma^2 +\frac{5 Ng^2}{192\sqrt{2} \pi}M^2=0,
      \label{eqGap1order}
\end{equation}
which is a cubic equation.
As a result, there are three solutions for this equations, i.e. there are three local minimums for the vacuum energy density. By construction of the Gribov-Zwanziger theory, in perturbative regime, the gap equation solution ensures that the functional integral of the gauge fields is taken in the region where the gauge field configurations are associated with the smallest eigenvalues of the Fadeev-Popov operator  \cite{Vandersickel:2012tz,Zwanziger:1988jt}. 
Therefore, although there are three possible solutions for the gap equation, we are guaranteeing that the gauge field configurations belonging to the integration domain correspond to those associated with the smallest eigenvalues of the Fadeev-Popov operator at the leading-order in $g^2$ \cite{Capri:2018ijg}. To get a global minimum, i.e. only one solution for the gap equation, the next order correction in $g^2$ may be used. However, it is not the goal of this paper. 

The discriminant of \eqref{eqGap1order} determines if these roots are complex or real. 
In consequence, if the discriminant of \eqref{eqGap1order} is positive,
\begin{equation}
\label{delta}
    \Delta= -
    \left(
    1215 \pi ^2 M^2-16 g^4 N^2
     \right)
    >0
  \end{equation}
there are three real roots, which means three real values for the Gibov parameter, $\gamma$.
On that account, $M$ has limited values to satisfy \eqref{delta},
\begin{equation}
\label{MFordeltaP}
  0<M<\frac{4 g^2 N}{9 \sqrt{15} \pi }.
  \end{equation}
 For values of $M$ bigger than $\frac{4 g^2 N}{9 \sqrt{15} \pi }$, $\gamma$ assumes complex values.
 Also, here, we have chosen positive values of Chern-Simons mass $M>0$.

By using the François Viète's formula, we write down all three real solutions for \eqref{eqGap1order}:
\begin{equation}
    \gamma_{t}=\frac{g^2 N}{18 \sqrt{2} \pi }+\frac{g^2 N}{9 \sqrt{2} \pi}\cos \PR{\frac{2 \pi  t}{3}-\frac{1}{3} \arccos\PC{1-\frac{1215 \pi ^2 M^2}{8 g^4 N^2}}},\label{gamma_tri}
\end{equation}
where $t=0,~1,~2$. 
Notice, if we take $M=0$, we recover the result from \cite{Canfora:2013zza} and only one solution for the gap equation, not three. 




If $\Delta=0$, we obtain
\begin{equation}\label{Max_M}
  M=\frac{4 g^2 N}{9 \sqrt{15} \pi }\;,  
\end{equation}
and the solutions of the gap equation are
\begin{equation}
    \gamma_t=\frac{g^2 N}{18 \sqrt{2} \pi }+\frac{g^2 N}{9 \sqrt{2} \pi } \sin \left(\frac{2 \pi  t}{3}+\frac{\pi }{6}\right),
\end{equation}
there are three roots, but two of them are similar $\gamma_0=\gamma_1$. We see that the value of the topological mass given by \eqref{Max_M} is the maximum allowed value of $M$ to obtain real roots for the gap equation. 

In \cite{Canfora:2013zza}, the weak coupling constant regime is given by the condition without restriction to Gribov horizon, that is
\[
M>\frac{Ng^2}{6\pi},
\]
eq. (8) in their paper, which is bigger than the maximum value of $M$ given by \eqref{Max_M}, $\frac{Ng^2}{6\pi}>\frac{4 g^2 N}{9 \sqrt{15} \pi }$. Therefore, in the case of real roots for the gap equation, the regime is always in the strong coupling regime, since $M$ is always smaller than $Ng^2/6\pi$. This result is different from the one in \cite{Canfora:2013zza}.

\begin{figure}
    \centering
    \begin{subfigure}{0.5\textwidth}
    \centering
     \includegraphics[width=\textwidth]{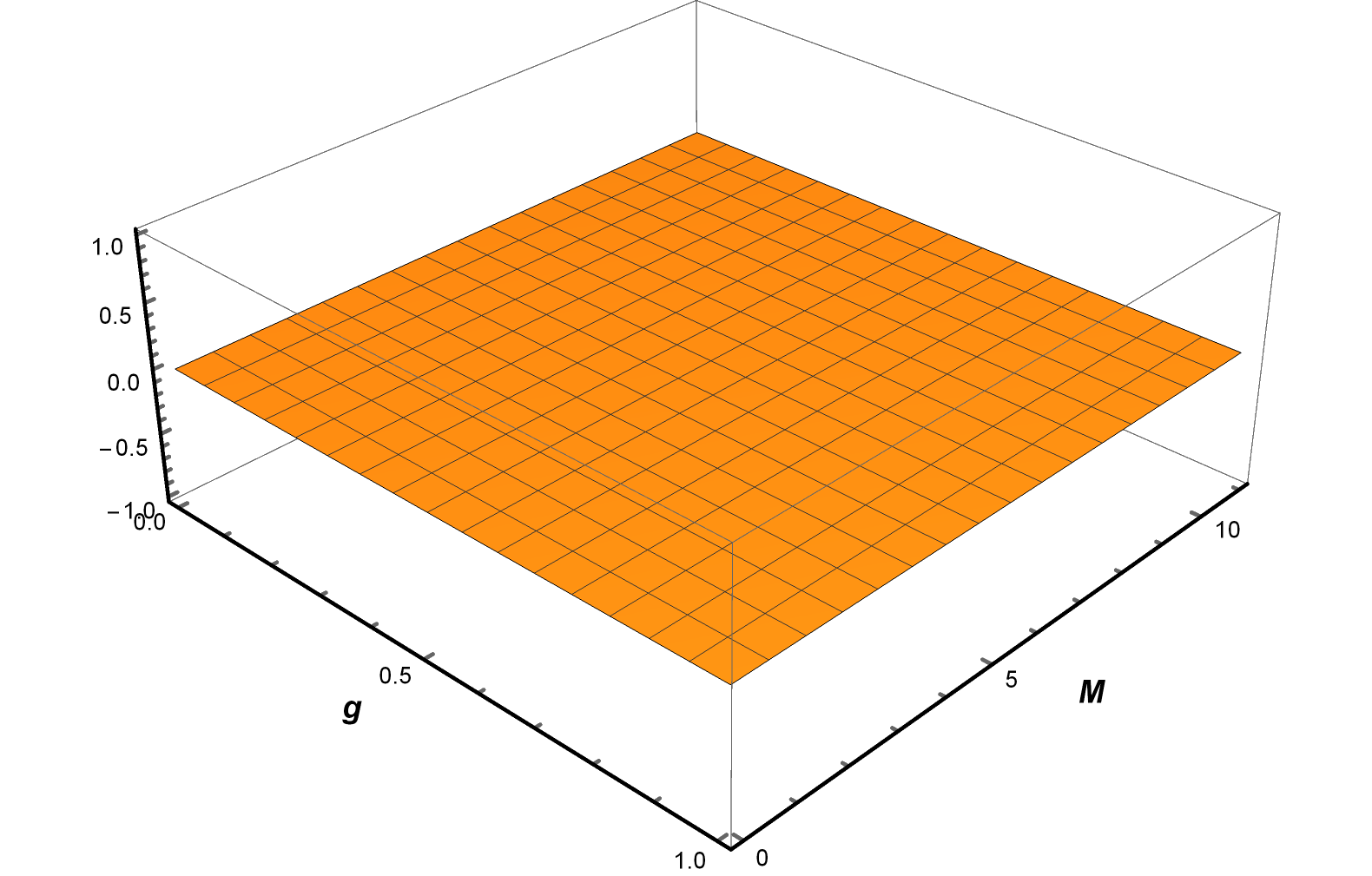}
     \caption{Im[$\mathfrak{G}_{1}(g,M)$]}
     \label{gc1}
     \end{subfigure}
     \hfill
    \begin{subfigure}{0.5\textwidth}
    \centering
    \includegraphics[width=\textwidth]{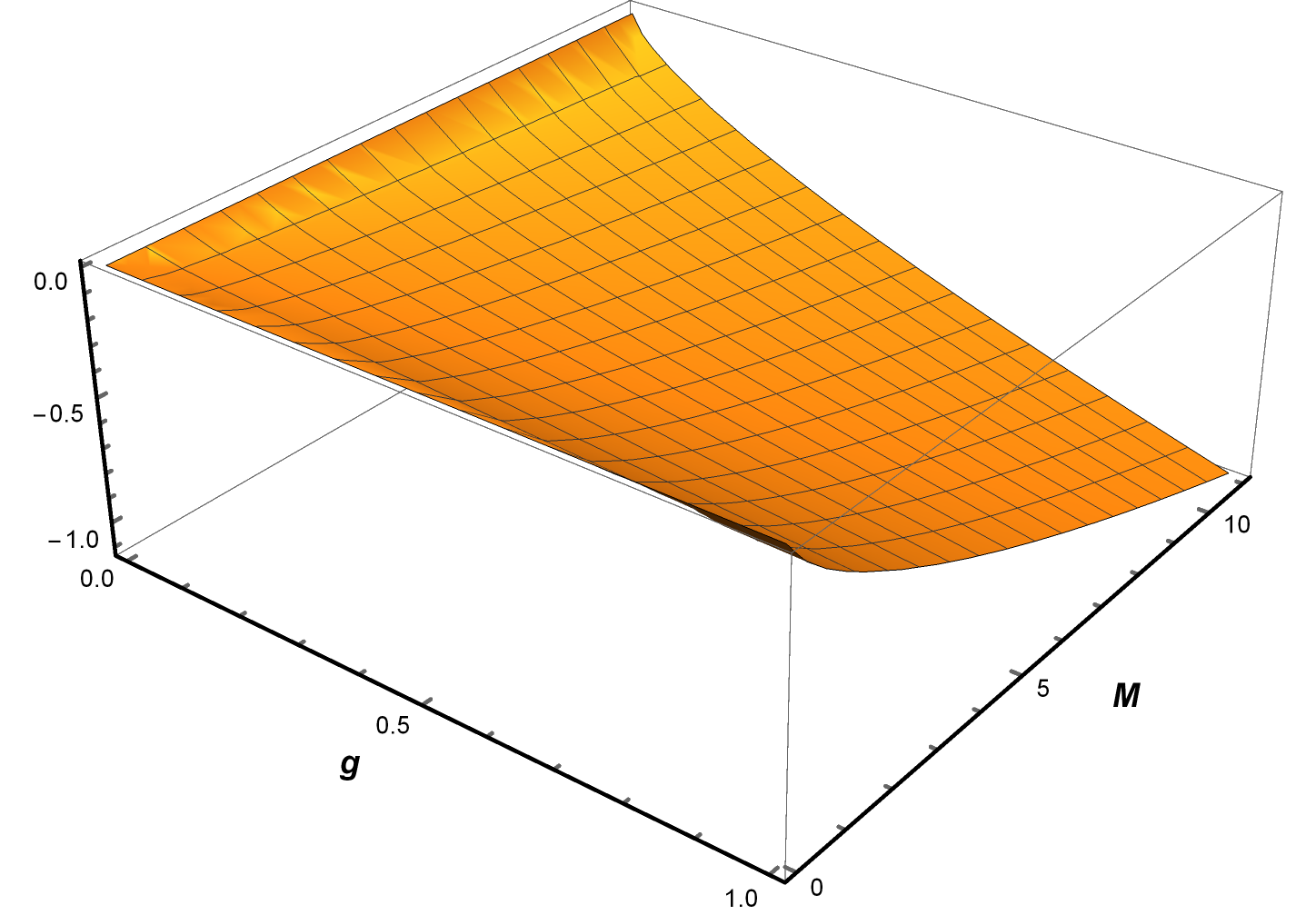}
    \caption{Im[$\mathfrak{G}_{2}(g,M)$]}
    \label{gc2}
    \end{subfigure}
     \hfill
    \begin{subfigure}{0.5\textwidth}
    \centering
    \includegraphics[width=\textwidth]{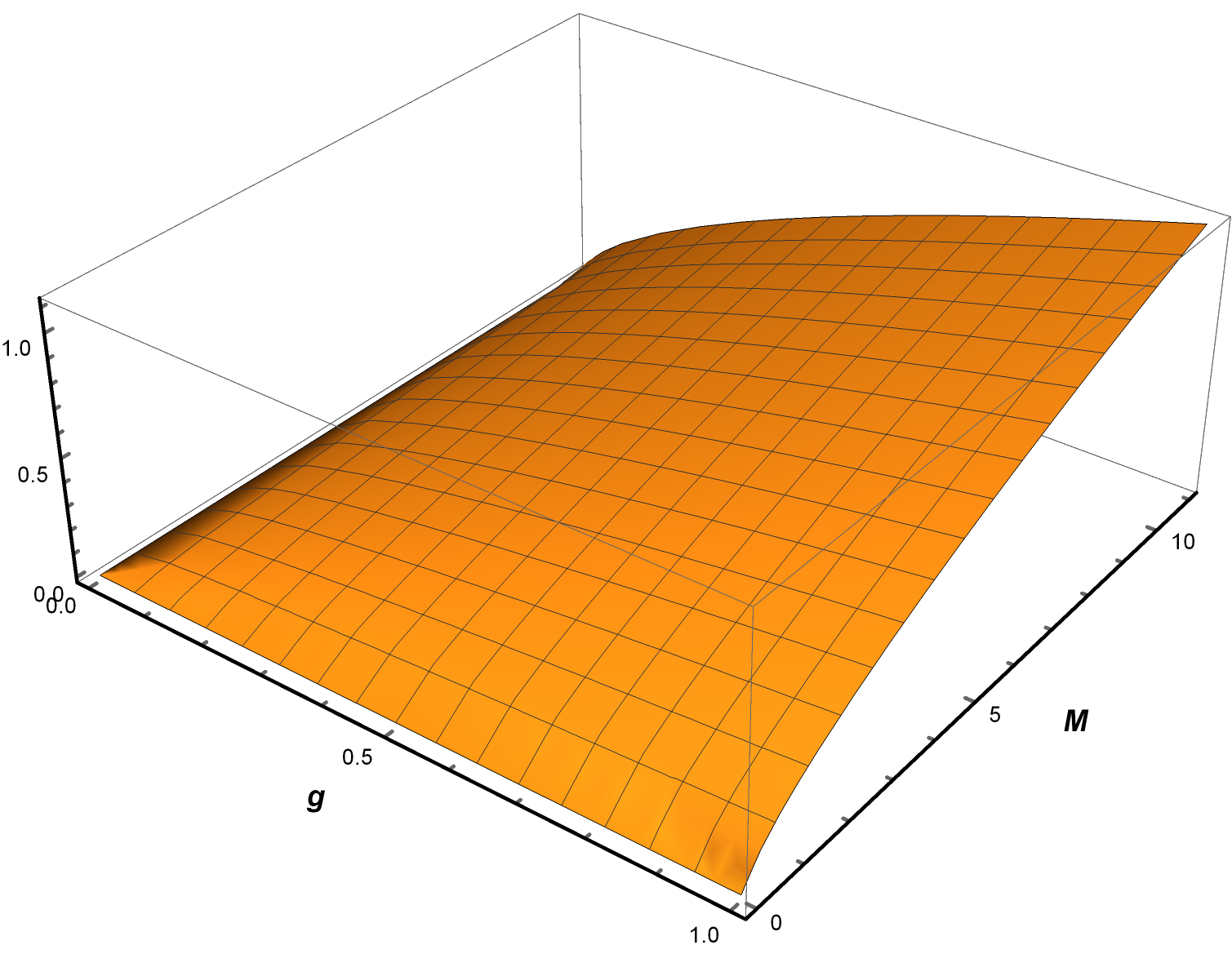}
    \caption{Im[$\mathfrak{G}_{3}(g,M)$]}
    \label{gc3}
    \end{subfigure}
    \caption{Plot of imaginary part of $\gamma$ when $\Delta<0$, $N=3$.}
    \label{gamma_solIm}
\end{figure}

Now, if $\Delta<0$, i.e.
\[
M>\frac{4 g^2 N}{9 \sqrt{15} \pi },
\]
there are one real root $\mathfrak{G}_{1}$, that is exactly equal the real root  $\gamma_2$ from the case $\Delta>0$, and two complex conjugate roots, $\mathfrak{G}_{2}$ and $\mathfrak{G}_{3}$. We do not show their expression in this paper, because they are complicated expressions in function of $g$ and $M$. Instead, we only plot their imaginary part, see Figure \ref{gamma_solIm}. Of course, the imaginary part of $\mathfrak{G}_{1}$ is null, since it is the real root.

After we have found the solutions for the gap equation, we can go back to the investigation about the gluon propagator poles from Section \ref{propagatorsec} and analyze the regime of theory for each Gribov parameter in the next section.
\section{The Regime of the Yang-Mills-Gribov-Zwanziger-Chern-Simons theory} \label{propagatorPoles}

In \cite{Canfora:2013zza}, the authors investigated the regime of the theory comparing $M$ and $g$, since, there, $\gamma\propto g^2$ and it is not a function of $M$. However, in our case, $\gamma$ also depends on $M$, see Section \ref{gapeq}, then it must be taken into account too. To do so, we separately investigate all solutions from \eqref{eqGap1order} using 
Wolfram Mathematica software.
Therefore, we go back to the analysis from Section \ref{propagatorsec}, where we presented the regime of the theory via the discriminant of the polynomial $P(k^2)$, \eqref{DeltaP}, and study all cases for $\gamma$. This examination is shown in Tables \ref{RegimebyRealRoots}  and \ref{RegimebyCompRoots}.

\begin{table}
\centering
\begin{tabular}{ |c| c|c|c|c| }
\hline
  & $\bf{\gamma_0}$ & $\bf{\gamma_1}$ & $\bf{\gamma_2}$ & \bf{$\mathfrak{G}_1$}\\
 \hline
 \multirow{3}{*}{\bf{Confinement}} & & & & \\
 & $0<M\leq\frac{4 g^2 N}{9 \sqrt{15} \pi }$ & $  \frac{\left(6 \sqrt{6}-5 \sqrt{2}\right) }{54 \sqrt[4]{3} \pi }N g^2<M\leq\frac{4 g^2 N}{9 \sqrt{15} \pi }$ & False & False\\ 
\bf{$\Delta_p>0$} &  & & &\\
 \hline
 \multirow{3}{*}{\bf{Deconfinement}} & & &  &\\
 & False & $ 0< M<\frac{\left(6 \sqrt{6}-5 \sqrt{2}\right) }{54 \sqrt[4]{3} \pi }N g^2$ & $ 0 <M\leq\frac{4 g^2 N}{9 \sqrt{15} \pi }$ & $ M>\frac{4 g^2 N}{9 \sqrt{15} \pi }$\\ 
\bf{$\Delta_p<0$} & & & & \\
  \hline
\end{tabular}
\caption{The regime of the theory defined by the real roots from the gap equation.}
\label{RegimebyRealRoots}
\end{table}

\begin{table}
\centering
\begin{tabular}{ |c| c|c|c|c| }
\hline
  & \bf{$\mathfrak{G}_2$} & \bf{$\mathfrak{G}_3$} \\
 \hline
 \multirow{3}{*}{\bf{Confinement}} & & \\
 & False & False\\ 
\bf{$\Delta_p>0$} &  &\\
 \hline
 \multirow{3}{*}{\bf{Deconfinement}} &  &\\
 & False & False\\ 
 \bf{$\Delta_p<0$}& & \\
  \hline
\end{tabular}
\caption{The regime of the theory defined by the complex conjugate roots from the gap equation.}
\label{RegimebyCompRoots}
\end{table}
 
By Table \ref{RegimebyRealRoots}, we see that $\gamma_2$ and $\mathfrak{G}_2$ do not describe the confinement phase. Therefore, gluons always behave as free particles. By the same table, we also realize there is no physical excitation when the Gribov parameter is given by $\gamma_0$. 
By Table \ref{RegimebyCompRoots}, we notice that the complex conjugate roots from the gap equation do not contribute to establishing the regime of the theory.

Therefore, we deduce that in order to define the regime of the theory, first, it is necessary to say in which Gribov parameter we are working with. For example, if we choose $\gamma_{2}$, all excitations in the theory are not confined even for $g$ big and $M$ small, which is opposite to the results from the study of \cite{Canfora:2013zza}. Also, 
for $\gamma_{0}$, 
all excitations in the theory are confined even for $g$ small and $M$ big, which diverges from the results in \cite{Canfora:2013zza}. 
As well, the only Gribov parameter that gives us both phases (confinement and deconfinement) is $\gamma_{1}$, see Table \ref{RegimebyRealRoots}.
However, by choosing the solution $\gamma_{1}$, the CS mass term assumes smaller values in the deconfinement phase than in the confinement phase, leading to an alternative conclusion than \cite{Canfora:2013zza}.  Furthermore, in the confinement defined by $\gamma_{1}$, see Table \ref{RegimebyRealRoots},
we see that when $g$ assumes large values, $M$ also can assume large values. Therefore, all excitations in the theory are confined for large values of $g$ and also for large values of $M$, which is a different result than \cite{Canfora:2013zza} again.

\begin{figure}
    \centering
    \begin{subfigure}{0.8\textwidth}
    \centering
    \includegraphics[width=\textwidth]{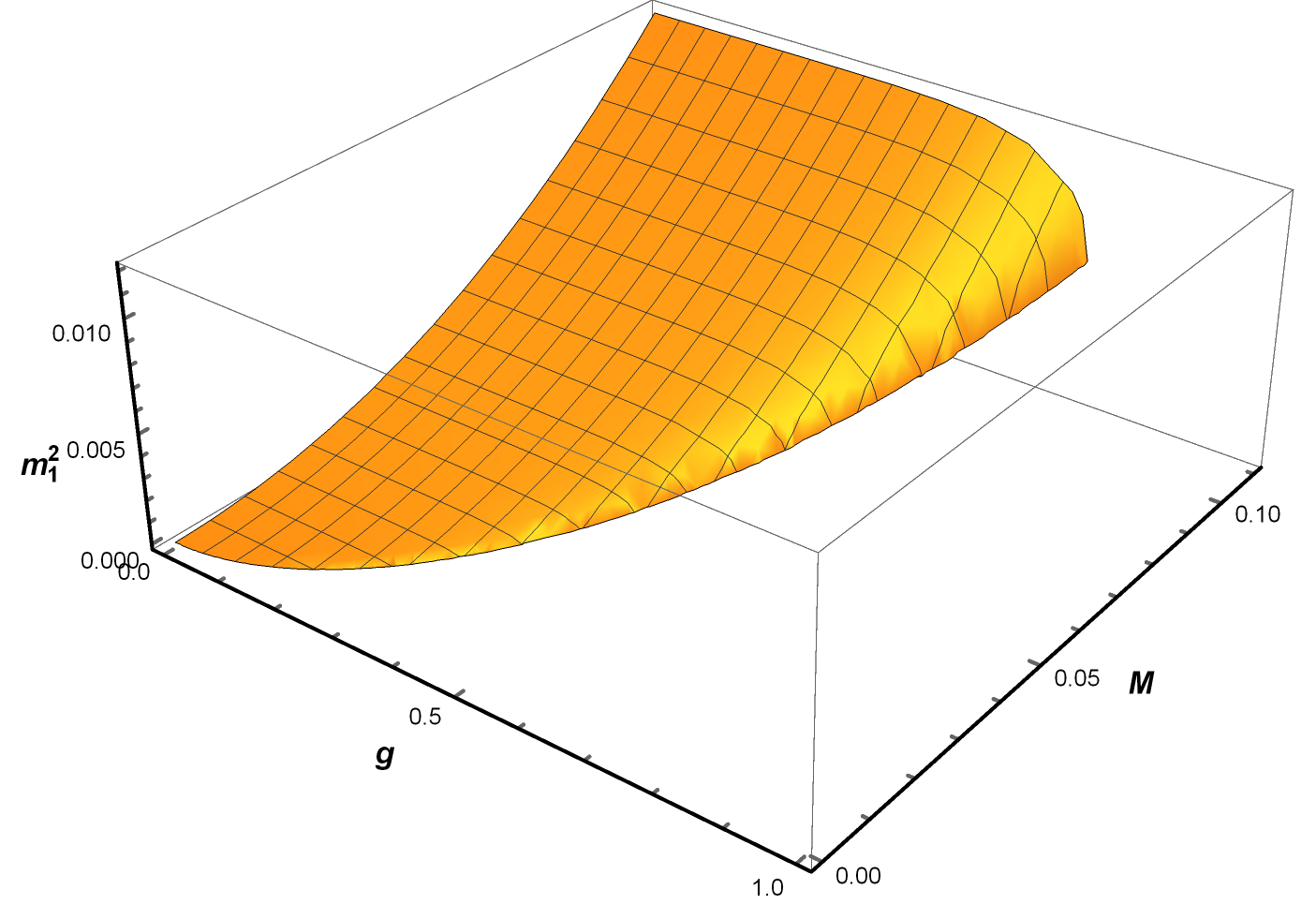}
    \caption{Zero order in $M^2$.}
    \end{subfigure}
     \hfill
    \begin{subfigure}{0.9\textwidth}
    \centering
     \includegraphics[width=\textwidth]{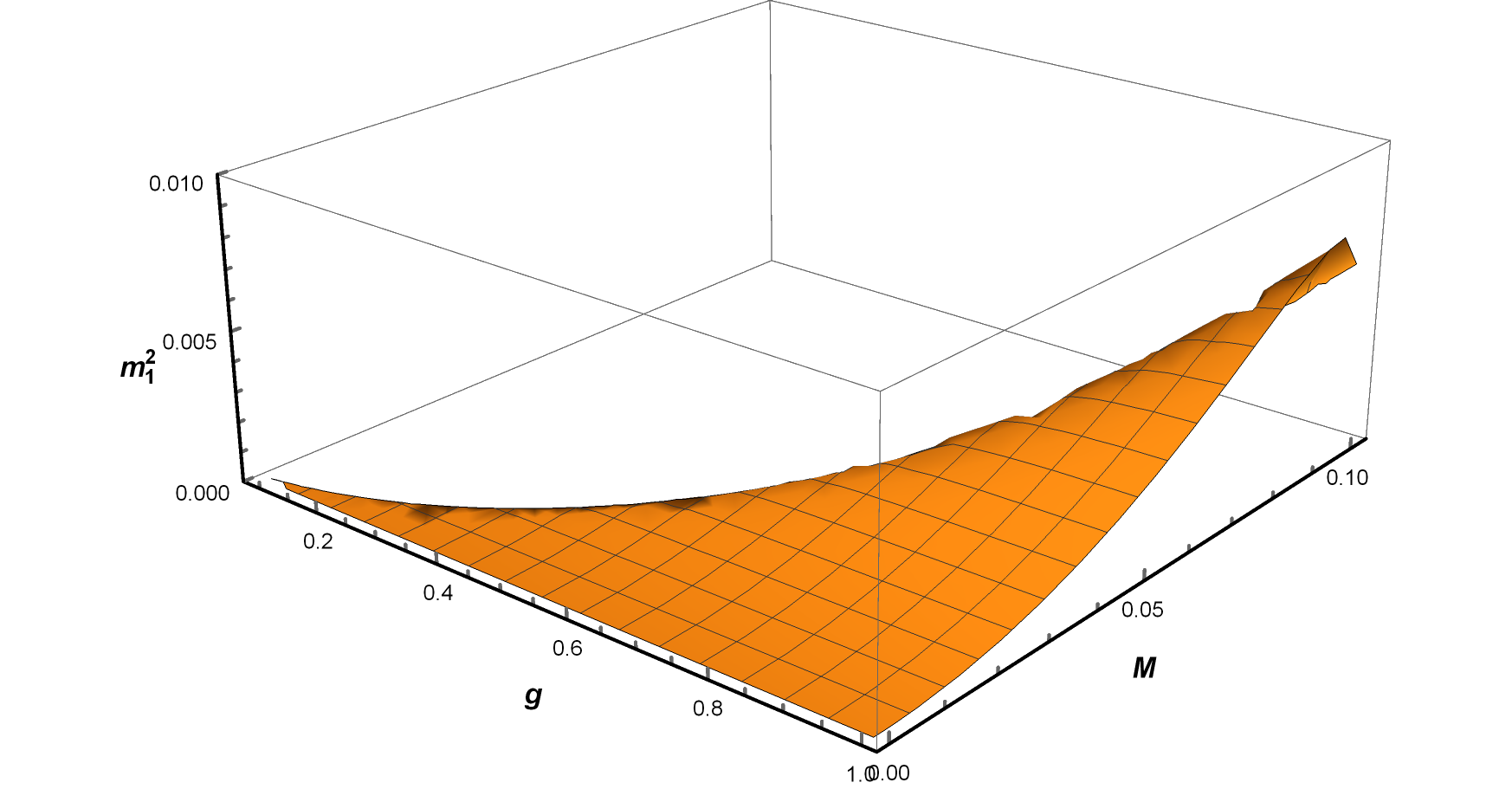}
      \caption{First order in $M^2$ by using $\gamma_1$.}
     \end{subfigure}
    \caption{$m_1^2$ in function of $M$ and $g$, $N=3$.}
    \label{pole2_sol}
\end{figure}

To visualize how the masses of the gluon propagator change by considering the first order in $M$, we plot $m_1^2$, the first pole from the polynomial polynomial $P(k^2)$, at zero order in $M$ and at first order in $M$ using $\gamma_1$, see Figure \ref{pole2_sol}. Clearly, they are very different from each other. Also, by this figure, we show that the poles of the gluon propagator are affected if the CS mass term is considered in the gap equation.

\section{Conclusion}
\label{conclusion}
In this paper, three solutions for 
the gap equation dependent on $(g,M)$ were presented, meaning that there are three local minimums for the vacuum energy density.
In \cite{Canfora:2013zza}, $M$ can assume all possible positive values. In our cases, this is not always true: $M$ assumes all possible positive values, if and only if we are working with $\gamma_{2}$ as shown in Table \ref{RegimebyRealRoots}.
However, if $\gamma_{2}$ is chosen, the confinement phase is not present, it means gluons are always free; in other words, only the deconfinement phase is observed in this case.
From Table \ref{RegimebyRealRoots}, we can also conclude that 
gluons are always confined when the Gribov parameter is given by $\gamma_{0}$. And, yes, for $g$ strong and $M$ small, there is confinement in agreement with the study of \cite{Canfora:2013zza}, but there is no deconfinement phase in the case of $\gamma_{0}$. The only Gribov parameter that gives us both phases (confinement and deconfinement) is $\gamma_{1}$.
In spite of that, with $\gamma_{1}$,
the CS mass term assumes smaller values than in the confinement phase, in contrast to the conclusion in \cite{Canfora:2013zza}. 

On that account, we have show that the regime of the theory is not only determined by comparing the coupling constant with the CS mass term, but also it depends on which Gribov parameter we are using.
In addition, in the case of $\gamma_{0}$, $\gamma_{1}$ and $\gamma_{1}$, the coupling constant is not small enough to get the weak-coupling regime, since $M$ is never bigger than $\frac{Ng^2}{6\pi}$. It means that there are always Gribov ambiguities and we are never in the perturbative regime. 
The weak-coupling regime is only reached with the solution given by $\mathfrak{G}_1$ -- a solution for the gap equation that only defines deconfinement phase, see Table \ref{RegimebyRealRoots}. 
Therefore, in this paper, we have shown that the gap equation solutions and the regime of the theory are evidently  affected by considering the first order in $M$ in the gap equation.

\section*{Acknowledgments}
This work was supported by Ministry of Science and Technology, Taiwan, grant number: MOST 109-2112-M-033
and 109-2811-M-033 -501.


\appendix
\section{Solving the gap equation}
\label{SolGapEg}
\begin{figure}
    \centering
    \includegraphics[width=0.8
    \textwidth]{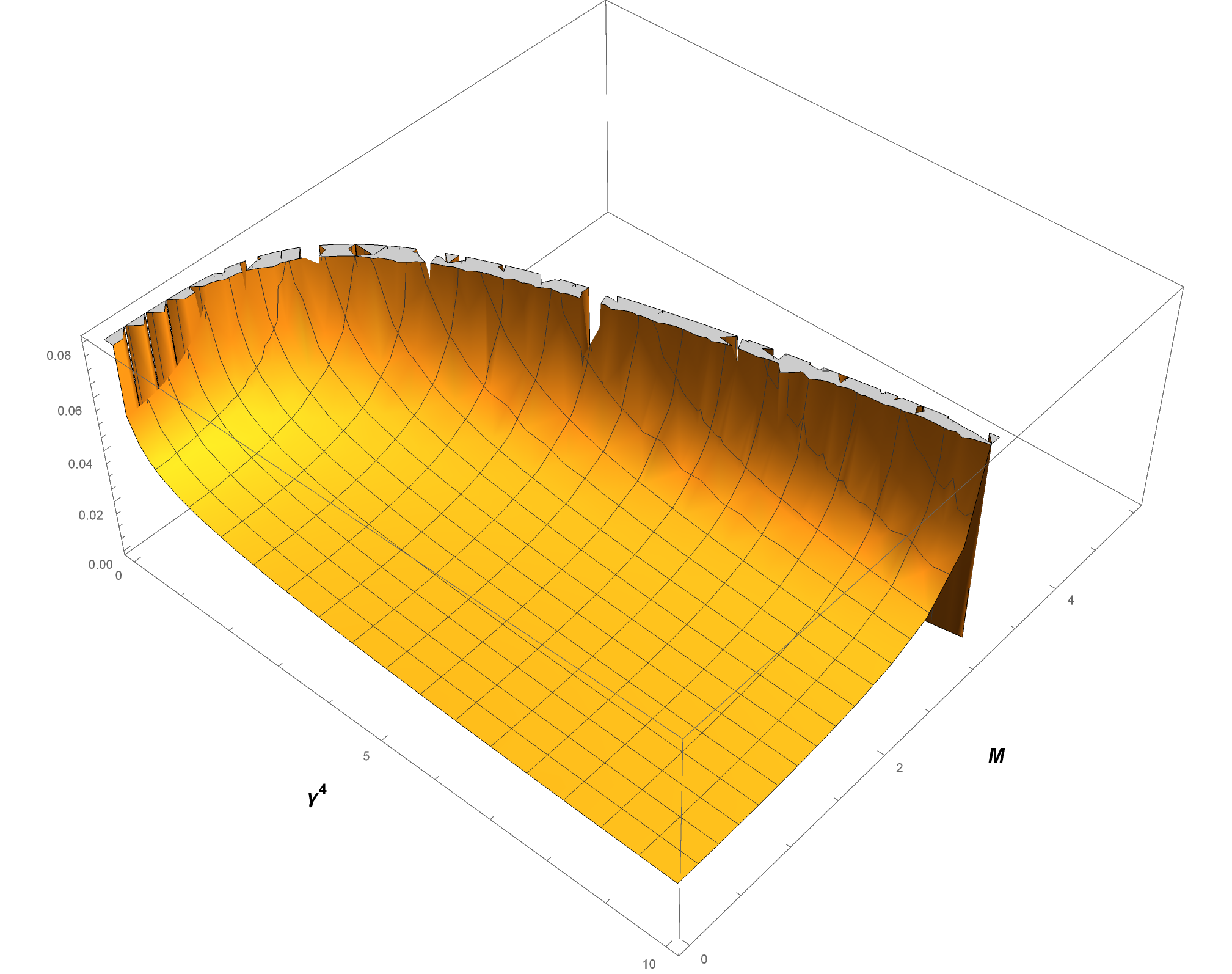}
    \caption{The left side from \eqref{gpequation3d}.}
    \label{leftSide}
\end{figure}

The integral on the left side of \eqref{gpequation3d} can be solved by the residues theorem. However, the solutions for the gap equation \eqref{gpequation3d} is only given by graphic analysis. The analysis of the equation will be done by comparing its left side shown in Figure \ref{leftSide}, with its right side. In the non-perturbative regime, i.e. in the infrared, $g^2$ is strong, the right side of \eqref{gpequation3d} is small. In Figure \ref{leftSide}, we can see that the left side from \eqref{gpequation3d} is also very small for $\gamma^4$ large and $M$ small. In the perturbative regime, i.e. when $g^2$ is small, the right side of \eqref{gpequation3d} is large. In the left side of the same equation, large values are reached when $M$ is big or $\gamma^4$ is small, what is consistent with theory, since $\gamma\propto \exp(-1/g^2)$.
In Figure 1, it can be observed that the left side of the gap equation  \eqref{gpequation3d} is not defined by all values of the Chern-Simons mass $M$, for $M>3$ in the plot, the function is not defined.

As we are working with Gribov formalism, we are only interested in the infrared (IR) regime, where Gribov copies show up. By the graphic analysis above, we see that the IR regime is given by a small $M$. Then, we can expand the left side of \eqref{gpequation3d} in $M^2$ using Taylor series. Hence, the gap equation can be written as follows
\begin{eqnarray}
&&
\int \frac{d^{3}k}{(2\pi)^{3}} \sum _{n=0}^{\infty } \frac{(-i)^{2 n} \left(k^{6 n} M^{2 n}\right)}{\left(\gamma ^4+k^4\right)^{2 n+1}}
~=~ 
\frac{3}{2Ng^2}
\end{eqnarray}
\begin{eqnarray}
&&
\sum _{n=0}^{\infty }(-i)^{2 n} \frac{4\pi}{(2\pi)^{3}} \int dk \frac{ k^{6 n} M^{2 n}}{\left(\gamma ^4+k^4\right)^{2 n+1}}k^2
~=~ 
\frac{3}{2Ng^2}\label{gapEqEnd}
\end{eqnarray}
In this paper, we only consider  the two first terms of \eqref{gapEqEnd}. This equation has been solved in zero order in \cite{Canfora:2013zza} as it was already said before. 

To solve \eqref{gapEqEnd}, we used the residue theorem:
\begin{equation}
\int_0^\infty f(x)dx=\frac12 \int_{-\infty}^{\infty} f(x)dx=\pi\ii \textnormal{Res}\PR{f(z)}_{z=z_0}=-\pi\ii \textnormal{Res}\PR{f(z)}_{z=- z_0}\\
\label{residue_general}
\end{equation}
where $z_0$ is the pole of the function $f(z)$.
In spherical coordinates, the zero order term from \eqref{gapEqEnd} is
\begin{equation}
\begin{split}
\frac{1}{2\pi^2}\int_0^\infty \frac{k^2}{k^4+\gamma^4}dk=
\frac{1 }{4 \pi\sqrt{2} \gamma}\label{zero_order_M},
\end{split}
\end{equation}
that is the result from \cite{Canfora:2013zza}. The first order term in $M^2$ of \eqref{gapEqEnd} is 
\begin{equation}\label{first_order_M}
    \begin{split}
       - \frac{4 \pi }{(2 \pi )^3}\int_{0}^{\infty}\frac{k^8 M^2}{\left(k^2-i \gamma ^2\right)^3 \left(k^2+i \gamma ^2\right)^3}dk=
     -\frac{5 M^2}{128 \sqrt{2} \pi  \gamma ^3}.
    \end{split}
\end{equation}
By replacing \eqref{zero_order_M} and \eqref{first_order_M} in \eqref{gapEqEnd}, we obtain

\begin{equation}
      \gamma^3-\frac{Ng^2}{6\sqrt{2} \pi}\gamma^2 +\frac{5 Ng^2}{192\sqrt{2} \pi}M^2=0.
      \label{eqGap1orderA}
\end{equation}

\appendix

\bibliographystyle{unsrt}
\bibliography{bib}

\end{document}